\input{aipcheck}

\documentclass[
    ,final            
  ]
  {aipproc}

\layoutstyle{8x11double}
 
\renewcommand\XFMtitleblock{%
  \XFMtitle
  \let\XFMoldpar\par
  \def\par{\XFMoldpar\def\par{\space 
             (on behalf of the VERITAS Collaboration)\XFMoldpar}}%
   \XFMauthors
   \let\par\XFMoldpar
   \XFMaddresses
   \XFMabstract
   \vspace{5pt}%
   \XFMkeywords
   \XFMclassification
 }

\begin{document}

\title{VERITAS Observations of the BL Lac Object 1ES 1218+304}


\author{Pascal Fortin}{
  address={Barnard College, Department of Physics and Astronomy, 3009 Broadway, New York, NY, 10027, USA}
}

\begin{abstract}
The VERITAS collaboration reports the detection of very-high-energy (VHE) gamma-ray emission from the high-frequency-peaked BL Lac object 1ES 1218+304 located at a redshift of $z=0.182$. A gamma-ray signal was detected with high statistical significance for the observations taken during several months in the 2006-2007 observing season. The photon spectrum between $\sim160$ GeV and $\sim1.8$ TeV is well described by a power law with an index of $\Gamma = 3.08 \pm 0.34_{stat} \pm 0.2_{sys}$. The integral flux above 200 GeV corresponds to $\sim6\%$ of that of the Crab Nebula. The light curve does not show any evidence for VHE flux variability. Using lower limits on the density of the extragalactic 
background light (EBL) in the near-IR to mid-IR we are able to limit the range of intrinsic energy spectra for 1ES 1218+304. We show that the intrinsic photon spectrum is harder than a power law with an index of $\Gamma = 2.32 \pm 0.37$. When including constraints from the spectra of 1ES 1101-232 and 1ES 0229+200, the spectrum of 1ES 1218+304 is likely to be harder than $\Gamma = 1.86 \pm 0.37$.
\end{abstract}

\maketitle


\section{Introduction}

In the TeV energy range, 19 BL Lac objects and 1 Radio Galaxy (M 87) have been established as emitters of TeV gamma rays. With the exception of the recently discovered low-frequency-peaked BL Lac (LBL) object BL Lacertae \cite{Albert:2007mz} and the intermediate-frequency-peaked BL Lac (IBL) object W Comae \cite{Acciari:2008p3184}, all TeV blazars are high-frequency-peaked BL Lac (HBL) objects. HBLs are a subclass of blazars characterized by a synchrotron peak at X-ray energies, unlike quasars that have higher luminosity and a synchrotron peak at optical/IR energies. 1ES 1218+304 is an X-ray-bright ($F_{1\mathrm{keV}} > 2\mu$Jy) HBL object located at a redshift $z=0.182$ \cite{Bade:1998ly}. Based on SED modeling and \textit{Beppo}SAX X-ray spectra, several of the HBLs were predicted to be TeV sources and several of them have indeed been detected at TeV energies \cite{Costamante:2002rt}. 1ES 1218+304 was predicted to be a good TeV candidate based on the position of its synchrotron peak at high energy and sufficient radio-to-optical flux. It was recently detected (at very high energies) by the MAGIC telescope \cite{Albert:2006yq}.

\section{VERITAS Observations}

The VERITAS observatory consists of an array of four 12-meter diameter imaging atmospheric Cherenkov telescopes (IACTs) located at the Fred Lawrence Whipple Observatory in southern Arizona \cite{T.-C.-Weekes:2002lr}. The array is sensitive to gamma rays in the energy range from 100 GeV to 30 TeV. The first two telescopes were operated in the stereoscopic observation mode from March 2006 and the third and fourth telescopes came online in December 2006 and April 2007, respectively. VERITAS observed 1ES 1218+304 from December 2006 to March 2007 using three telescopes. The data were taken in wobble mode where the source is offset from the center of the field of view by $0.5^\circ$ and the background is measured directly from different regions in the same field of view but away from the source region. After removing data taken under poor sky conditions or affected by various detector problems, we were left with a total observation time of 17.4 hours covering a range in zenith angle from $2^\circ$ to $35^\circ$, with an average zenith angle of $14^\circ$. 

\section{Analysis and results}

The standard analysis tools were used to process the data. After calculating the standard Hillas parameters, images with an integrated charge less than 400 digital counts\footnote{An integrated charge of 400 digital counts calculated using a 10 ns integration window corresponds to $\sim$75 photoelectrons.}  or with a distance from the center of the camera larger than $1.2^\circ$ were rejected. The shower direction in the field of view and the impact parameter of the shower core were calculated using stereoscopic techniques \cite{Hofmann:1999lr,Krawczynski:2006lr}. The background was estimated using the \textit{reflected-region} and \textit{ring background} methods as described in \cite{Berge:2007p689}. A set of scaled cuts on the width and length parameters were used to identify gamma-ray events in the data. An extensive set of Monte Carlo simulations was used to generate lookup tables and calculate the effective area of the detector as a function of zenith angle and gamma-ray energy.

\begin{figure}
\includegraphics[height=.3\textheight]{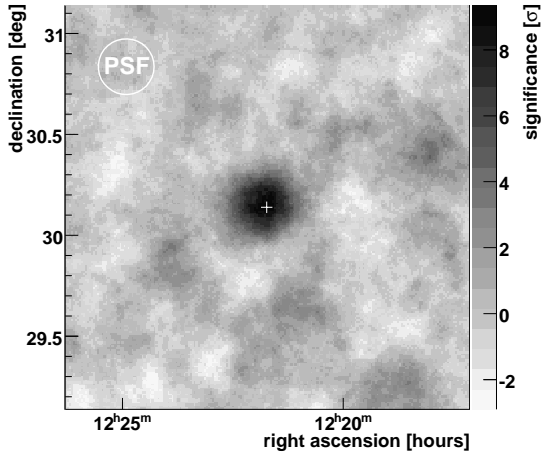}
\caption{Significance map of the region around 1ES 1218+304. The white cross indicates the position of the radio source. The white circle in the upper left corner shows the angular resolution of VERITAS (PSF). The ring background model was used to estimate the background and the significances were calculated using equation 17 in \citet{Li:1983lr}.}
\label{fig1}
\end{figure}

Figure \ref{fig1} shows a two-dimensional sky map of significances for a region centered on the radio coordinates of 1ES 1218+304 ($12^{h}21^{m}21.9^{s}, +30^{\circ}10'37''$ J2000). The excess is compatible with that expected from a point source. Fitting a two-dimensional Gaussian distribution to the uncorrelated excess map yields a peak position in good agreement with the radio coordinates.

\begin{figure}
\includegraphics[height=.3\textheight]{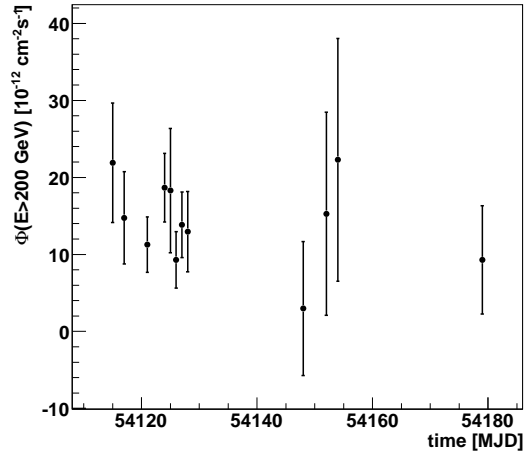}
\caption{Light curve of the integral photon flux above 200 GeV for the source 1ES 1218+304. Each point corresponds to the average daily flux and assumes a spectral shape of dN/dE $\propto$ E$^{-\Gamma}$ with $\Gamma = 3.08$. The error bars represent the statistical uncertainty.}\label{fig2}
\end{figure}

\begin{figure}
\includegraphics[height=.3\textheight]{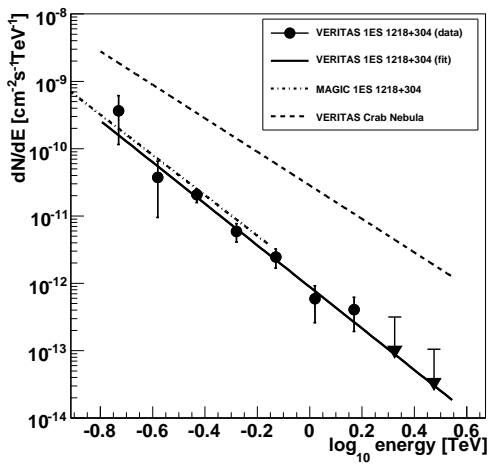}
\caption{Differential energy spectrum of VHE photons above 160 GeV for 1ES 1218+304. The markers indicate measured data points and the continuous line is a power-law fit. Downward pointing arrows correspond to upper flux limits (99\% probability, \citet{helene}) for bins with significances below two standard deviations. The dot-dash line shows the differential energy spectrum of 1ES 1218+304 measured by MAGIC \citep{Albert:2006yq} and the dash line shows the differential energy spectrum of the Crab Nebula measured by VERITAS (September to November 2006) for comparison.}
\label{fig3}
\end{figure}

Figure \ref{fig2} shows the light curve of the integral flux above 200 GeV for the months of January, February, and March 2007. The average integral flux was calculated for each day assuming a spectral shape of dN/dE $\propto$ E$^{-\Gamma}$ with $\Gamma = 3.08$. A statistical test ($\chi^2$/dof = 6.7/11) indicates that no statistically significant variability was detected in the data. Figure \ref{fig3} shows the time-average differential energy spectrum for  gamma-ray energies between 160 GeV and 1.8 TeV. The shape is consistent with a power law ($\chi^{2}/\textrm{dof} = 2.1/5$) $\mathrm{dN/dE} = \mathrm{C\times(E/0.5TeV)}^{-\Gamma}$ with a photon index $\Gamma = 3.08 \pm 0.34_{stat} \pm 0.2_{sys}$ and a flux normalization constant $\mathrm{C} = (7.5 \pm 1.1_{stat} \pm 1.5_{sys}) \times 10^{-12} \mathrm{cm^{-2}s^{-1}TeV^{-1}}$. The integral flux is $\Phi(\textrm{E} > 200 \textrm{ GeV}) = (12.2 \pm 2.6) \times 10^{-12} \mathrm{cm^{-2}s^{-1}}$ which corresponds to $\sim6$\% of the flux of  the Crab Nebula above the same threshold.

\section{Discussion and conclusions}

Due to the relatively large redshift of 1ES 1218+304 ($z=0.182$) we expect significant attenuation from the interaction of high-energy gamma rays with low-energy photons from the EBL. TeV gamma-rays from extragalactic sources are expected to pair produce with optical/IR low-energy photons as they travel through intergalactic space, leading to a cutoff in the measured gamma-ray spectrum. Intrinsic spectra of blazars are not known a priori. Nevertheless, it is possible to use the EBL lower limits from galaxy counts to restrict the range of EBL scenarios / models and discern a corresponding range of intrinsic blazar spectra.

Here we present possible intrinsic energy spectra of 1ES 1218+304 that are compatible with EBL lower limits from galaxy counts. Furthermore, we broaden our study and include the energy spectra of 1ES 1101-232 and 1ES 0229+200 as measured by H.E.S.S. (see \cite{Aharonian:2006p1822,Aharonian:2007p1477} respectively) to reach a conclusion about the intrinsic energy spectra of this small sample of TeV blazars. The combination of energy spectra in the sub-TeV to TeV waveband at large redshfift (1ES 1102-232, 1ES 1218+304) with energy spectra in the multi-TeV regime at moderate redshift (1ES 0229+200) provides additional sensitivity to EBL spectra and their relative intensities in the near-IR and mid-IR. We have considerd a large variety of EBL scenarios with different spectral indices and shapes. A convenient parameterization of the EBL scenarios, provided by \cite{Dwek:2005ve}, is used in this study for providing a limit to the hardness of the blazar spectrum of 1ES 1218+304 as it provides a wide range of EBL spectra with different near-IR to mid-IR ratios.

Table \ref{tbl-1} shows the intrinsic energy spectra of all 3 blazars for a range of EBL scenarios. The ones that are still compatible with lower limits from galaxy counts represent the softest possible gamma-ray spectra. When considering 1ES 1218+304 by itself, the softest intrinsic spectrum is described by a power law with $\Gamma=2.32\pm0.37$ and is derived from a scaled version of the LHL scenario (LHL0.76). However, when applying this analysis to previously reported blazar spectra 1ES 1101-232 and 1ES 0229+200, the LHL0.76 scenario would require an extremely hard intrinsic spectrum for 1ES 0229+200. Therefore, in order to provide  a limit to the hardness of the blazar spectra based on these three sources we search for the softest possible intrinsic spectrum. As can be seen in Table \ref{tbl-1}, a search for the softest possible blazar spectrum among the sample of three blazars yield an EBL scenario (AHA x 0.45) that still requires the spectrum of 1ES 1218+304 to be as hard as $\Gamma=1.86\pm0.37$. All other EBL scenarios yield harder spectra for one of these three blazars.

\begin{table}[!t]
\begin{tabular}{cccc}
\hline
\tablehead{1}{c}{b}{Scenario} &
\tablehead{3}{c}{b}{$\Gamma_{source}$} \\
\cline{2-4} &
\tablehead{1}{c}{b}{1ES~1101-232} &
\tablehead{1}{c}{b}{1ES~1218+304} &
\tablehead{1}{c}{b}{1ES~0229+200}\\
\hline
  AHA1.0      &   $\rm  0.44 \pm 0.20  $  &  $ \rm  0.35 \pm 0.40   $  & $ \rm  2.35 \pm 0.14$ \\
  AHA0.55      &   $\rm 1.54 \pm 0.20  $  &  $ \rm  1.59 \pm 0.37   $   & $ \rm  2.41 \pm 0.14$  \\
  AHA0.45      &   $\rm 1.78 \pm 0.20  $  &  $ \rm  1.86 \pm 0.37   $   & $ \rm  2.43 \pm 0.13   $   \\
  HHH       &   $\rm -0.67 \pm 0.12  $  &  $ \rm  -0.72 \pm 0.29   $ & $ \rm  0.90 \pm 0.17   $  \\
  LLH$^{*}$       &   $\rm 2.01 \pm 0.22  $  &  $ \rm  2.07 \pm 0.35   $  & $ \rm  2.12 \pm 0.20   $    \\
  LHL       &   $\rm 2.04 \pm 0.20  $  &  $ \rm  2.08 \pm 0.39   $  & $ \rm  0.94 \pm 0.32 $     \\
  LHL0.70$^{*}$   &   $\rm 2.32 \pm 0.21  $  &  $ \rm  2.43 \pm 0.37   $  & $ \rm  1.43 \pm 0.29   $      \\
  LHL0.76   &   $\rm 2.23 \pm 0.21  $  &  $ \rm  2.32 \pm 0.37   $  & $ \rm  1.30 \pm 0.29   $      \\
  LHL0.82   &   $\rm 2.18 \pm 0.21  $  &  $ \rm  2.26 \pm 0.38   $  & $ \rm  1.20 \pm 0.30   $      \\
  MHL0.70  &   $\rm 1.26 \pm 0.19  $  &  $ \rm  1.34 \pm 0.36   $  &   $ \rm  1.35 \pm 0.21  $      \\
  MHL0.55  &   $\rm 1.61 \pm 0.19  $  &  $ \rm  1.73 \pm 0.35   $  &   $ \rm  1.59 \pm 0.20  $     \\
  LLL$^{*}$       &   $\rm 2.06 \pm 0.16  $  &  $ \rm  2.20 \pm 0.34   $    &  $ \rm  2.11 \pm 0.20   $      \\
\hline
\end{tabular}
\caption{The absorption-corrected spectral indices ($\rm dN/dE \propto E^{- \Gamma_{source}}$) of 1ES~1101-232,  1ES~1218+304, and 1ES~0229+200 from a small sample of all EBL scenarios considered. EBL scenarios that fall below the lower limits from galaxy counts are marked with an asterisk.  Scaled Aharonian scenarios (AHA$x.y$)  are taken from \cite{Aharonian:2006p1822}. HHH (High near-IR, mid-IR, far-IR)m MHL (Medium near-IR, High mid-IR, Low far-IR), etc, are taken from \cite{Dwek:2005ve}.}
\label{tbl-1}
\end{table}

In conclusion, these results indicate that blazar spectra as evidenced by the two most distant blazars, 1ES 1101-232 and 1ES 1218+304,  are hard. These values are still within the acceptable range predicted for shock acceleration in blazars (see, e.g. \cite{Stecker:2007p2209}), however they are also close to a limit ($\Gamma = 1.5$) that was previously suggested by \cite{Aharonian:2006p1822}. The results presented here from VERITAS observations confirm with high statistical significance the MAGIC discovery of the HBL object 1ES~1218+304 as a source of VHE gamma rays. The detection of 1ES~1218+304 together with the other blazars studied by VERITAS \cite{Cogan:2009} give valuable insights into the physics of these sources and the EBL.


\begin{theacknowledgments}
This research is supported by grants from the U.S. Department of Energy, the U.S. National Science Foundation, and the Smithsonian Institution, by NSERC in Canada, by PPARC in the UK and Science Foundation Ireland.
\end{theacknowledgments}



\bibliographystyle{aipproc}   




\end{document}